\documentclass[a4paper]{article}
\begin{document}

\title{Supersymmetry with composite bosons}
\author{Alejandro Rivero\thanks{Zaragoza University at Teruel.  
           {\tt arivero@unizar.es}}}
\maketitle

\begin{abstract}
We note that hadronic susy (empirical quark-diquark) symmetry can be
expanded into the lepton sector, and that for three generations the counting
of degrees of freedom is the one we need to build charged supermultiplets. For
this to cure hierarchy, Higgs modeling becomes restricted.
\end{abstract}

Quark-Diquark supersymmetry has been an ongoing topic since the early works
 of Lichtenberg\cite{lich} and Miyazawa\cite{miya:old} in 1968. At these 
times the second generation
 of quarks and leptons was just coming out of the closet.

Now we have three (and no more) generations of quarks and leptons, and one
 of them, the top, is unable to form mesons, nor diquarks. In this situation 
hadronic supersymmetry presents a curious phenomena: it provides the bosonic
 degrees of freedom needed to cancel fermion loops. Of if you prefer, to build
 supersymmetric multiplets.

To be precise, we can form 15 different diquarks (and its corresponding 
antiquarks). Of them, we have 6 with charge -2/3, 6 with charge +1/3, and 3 
spurious -we hope- ones with charge +4/3.
Leaving aside for a moment the later ones, we can start to be intrigued 
about the coincidence.

If, departing from the venerable lines of pure hadronic thinking, we turn
now our view to the leptons, our intrigue grows to surprise as the diquarks
turn mesons: they provide us 6 charged +1 degrees of freedom, ready to fit
with the charged leptons, and of course the corresponding 6 antiparticles to
fit charged antileptons. We have also 13 neutral mesons, and some of them
should be susy to neutrinos before the application of the seesaw mechanism.
How many of them we can not tell, because it is model dependent. Also, group 
theoretical arguments could be used to decrease the available neutral
degrees.

To resume: for three generations of quarks and charged leptons, we find that
diquarks (and mesons) provide just the exact number of scalars needed to build
the corresponding charged supermultiplets. There are also extra 4/3 scalars, 
and a plethora of neutral ones, but the interesting point here is that 
we have a triple coincidence, for up, down, and electron types.

One could check that this coincidence needs a minimum of three generations, and
it is unique if only one quark lacks mesons. Consider N generations, with D
down-type quarks able to build mesons, and U up type quarks in the same
condition. Asking for coincidence, we obtain two equations

\begin{eqnarray*}
D \  U  &=& 2 N \\
D(D+1)/2 &=& 2N   
\end{eqnarray*}

The minimal solution needs N=3, and an extra condition, such as $U-D=1$, 
fixes this solution to be unique.

The extension of quark-diquark supersymmetry to a fundamental symmetry 
including leptons solves another amusing phenomenological question: what
the heck are muon and tau doing at the same energy levels that hadronic
quantities? Up to now, the only valid answer was to relate them to
the masses of strange of bottom, running down from a GUT scale. The relation
with down-kind quarks remains, because the mesons supersymmetric to 
charged leptons are the ones having the same composition that the diquarks
supersymmetric to down-kind quarks.

Marginally, lets point out that both down-like quarks and charged leptons are
known to fulfill Koide's relationship, a relation between three generations
of composite particles that amazingly works for leptons. Here we have a way
to explain this relationship: charged leptons are not composite, but they 
are supersymmetric to composite particles.

What about the hierarchy problem? In the standard model, diquarks (mesons) 
are just higher loop quark (lepton) corrections to the Higgs self energy. In
order for them to be able to cancel the one loop quark (or lepton) 
corrections, the Higgs structure must be such that it couples with diquarks
at the same coupling intensity than it couples with quarks. It does not seem
an impossible task to fulfill, e.g. with composite Higgs models. If 
additionally the interaction with double-up diquarks (the 4/3 ones) is not
favored, then we have got to get rid of our three spurious degrees of 
freedom. 

The possibility to control the hierarchy problem is insinuated by Miyazawa in
a short 1983 note \cite{Miyazawa:1983rz}, but leptons are not worked out,
nor the full third generation.
 
Finally, how has this supersymmetry been broken? One conjecture, based in the
preservation of Koide's formula for leptons, could be that the symmetry has
been broken in the meson side, and that for some value of the
strong coupling constant the supersymmetry could be restored. On other side
the top quark is very far from their candidate partners, and we could suspect
that the same mechanism breaking electroweak symmetry also breaks symmetry.

We should investigate new ways to break supersymmetry. As a suggestion, let
me point out that the algebra of functions over superspace has not been yet 
worked inside Connes formalism, and we know that this formalism has 
proved strong enough to produce the usual standard model and higgs. Still, the
most naive connection, that susy restoration happens when the discrete
dirac operator becomes infinitesimal, works in opposite sense to our desires
because an
infinite Higgs mass could imply an infinitely massive top quark.
 
\begin{table}
\begin{center}
\begin{tabular}{c|c|c}
4/3  & 1/3  & -2/3 \\
\hline
    &     & bb \\
\hline 
    & cb  & sb \\
\hline
    & ub  & db \\
\hline
cc  & cs  & ss \\
\hline
    & us  &    \\
uc  &     & ds \\
    & dc  &    \\
\hline
uu  & ud  & dd \\
\hline
\end{tabular}
\end{center}
\caption{All the possible diquark pairings, ordered according electric charge. Note
that in the 1/3 and -2/3 columns there are the exact number we need to form
susy multiplets with three generations of (anti)quarks. The
antiparticles follow the same pattern.} 
\end{table}

\begin{table}
\begin{center}
\begin{tabular}{c|c|c}
 +1  &   0    &  -1    \\
\hline     
     &  bb    &    \\
\hline
cb   & sb bs  & bc \\
\hline
ub   & db bd  & bu \\

\hline
cs   &   ss   & sc \\
     &   cc   &     \\
     
\hline         
us   &        & su  \\
     & ds sd  &     \\
     & cu uc  &     \\
dc   &        & dc  \\
\hline              
ud   &  dd    & du  \\
     &  uu    &     \\
\hline
\end{tabular}
\end{center}
\caption{all the possible quark/antiquark (meson) pairings, ordered according electric
charge. The charged ones are exactly the needed number to from three generations
of susy multiplets with the charged leptons}
\end{table}

\end{document}